\UseRawInputEncoding 

\documentclass[11pt, thmsa]{article}  
\usepackage{amsmath}
\usepackage{amsthm}
\usepackage{graphicx}
\usepackage{amssymb}

\newtheorem{mydef}{Definition}
\newtheorem{assu}{Assumption}
\newtheorem{prin}{Principle}
\newtheorem{condition}{Condition}

\newtheorem{remark}{Remark}[section]
\newtheorem{proposition}{Proposition}[section]
 \usepackage{mathptmx}      
\begin{document}

\title{A New Derivation of the Propagator's Path Integral for Spinless Elementary Particles}

\author{Domenico Napoletani\thanks{University Honors Program and Institute for Quantum Studies, Chapman University, Orange, CA. Email: napoleta@chapman.edu}, Daniele C. Struppa\thanks{ The Donald Bren Presidential Chair in Mathematics, Chapman University, Orange, CA. Email: struppa@chapman.edu.}}

\date{}

\maketitle

\begin{abstract}
\noindent
We introduce a notion of isolated units, elementary particles or more general physical phenomena that do not significantly affect their surrounding environment, and we build a primitive ontology to describe their evolution and interaction. We further introduce a notion of indistinguishability of distinct spacetime paths of a unit, for which the evolution of the state variables of the unit is the same, and a generalization of the equivalence principle based on indistinguishability. Under a time invertibility condition on the whole set of indistinguishable paths of a unit, we show that the quantization of motion of spinless elementary particles in a general potential field can be derived in this framework, in the limiting case of weak fields and low velocities.
Extrapolating this approach to include weak relativistic effects, we explore possible experimental consequences.

\end{abstract}

\newpage

\section{Introduction}

In this paper we build a framework where physical events are defined with respect to the evolutions of the state variables of distinct {\it isolated units} (what we will call {\it the experiences} of the units), and the intermittent matching of their experiences during interaction. These units will be formally defined in Section 2 as elementary particles, or more general physical phenomena, whenever they do not significantly affect their surrounding environment. 
We will demonstrate that some basic additional assumptions on the notion of experience of an isolated unit are sufficient to derive the quantum mechanical propagator for spinless elementary particles in the limiting case of weak fields and low velocities. 

The theoretical framework we propose is a self-consistent primitive ontology \cite{allori} of isolated units, in the sense that the theory affirms the reality of its constituent objects and of the corresponding experiences. Accordingly, since in isolation a unit is the only phenomenon validating the reality of a specific evolution of its state variables, the experience of an isolated unit is a physically valid local representation of reality, until the unit is forced to match its experience in interaction with another unit.

Note that the primitive ontology of isolated units does not rely on any preferred role of conscious observers (or even less, of conscious phenomena) in determining the outcome of physical processes. Indeed, the notion of experience of a unit and the notion of matching of experiences that we propose in Section 2 simply assume that to properly understand physical evolution and interaction of phenomena we must first consider the restricted information available to the phenomena directly involved. The framework of units of experience can then be seen as a contribution to the understanding of the role of information and computation in the foundations of quantum systems \cite{zell,hooft}.

Additionally, the framework we develop is relational, since it relies on matching distinct experiences, and it allows for multiple and compatible possibilities of motion of a unit to be meaningfully considered real physical quantities, as we will show in Section 2.2. These general characteristics are shared respectively with relational quantum mechanics \cite{rovelli} and with the transactional interpretation of quantum mechanics \cite{cramer,kastner}. We also note that the emphasis on the perspective of distinct phenomena has been already suggested in a different form in \cite{smolin}, in the broader context of the real ensemble interpretation of quantum mechanics \cite{smolin2}.  In this approach, to every element of a physical system, it is associated a ``view'' encoding the knowledge of the system according to that specific element.

Broadly speaking, the paper is divided in two parts. The longer Section 2 is is concerned with the elements of the primitive ontology of isolated units, that establishes the objects of the theory and their properties. We introduce the notions of isolated unit and of experience of a unit and we list their basic properties, including the assumption of an internal time state, which we then relate to the de Broglie hypothesis on the existence of an internal clock frequency, first postulated in \cite[Chapter 1]{deBroglie1} and expanded in its scope in recent years \cite{dolce1,dolce2,kastner2,mueller,refDeBroglie1}. We further introduce a notion of indistinguishability of distinct spacetime paths of a unit, for which the evolution of its state variables is the same, and a generalization of the equivalence principle based on indistinguishability.

Starting with Section 3, we develop the mathematical and physical apparatus implicit in the axioms and principles from Section 2. More particularly, the main result of Section 3 is the construction of a relativistically invariant path integral propagator for simple units, i.e units whose state variables do not change in time. This construction is dependent on a time invertibility condition on the whole set of indistinguishable paths of a unit and on identifying the frames of reference that can be meaningfully defined for isolated units. In Section 4 we show that the propagator for simple units approximates the quantum mechanical propagator of a spinless elementary particle in a potential field, in the limiting case of weak potentials and low velocities. Finally, in Section 5 we extrapolate the results of Section 4 to include weak relativistic effects on the propagator of a charged simple unit and we suggest the outline of an experimental setting that could detect small deviations from the predictions of standard quantum mechanics, under restrictive conditions on the extremal trajectory of the charged unit to minimize field radiation.

\section{The Evolution of Isolated Phenomena}

\subsection{Isolated Units and Internal Clock Frequency}
We now formally define isolated units, and we introduce the key properties necessary to understand their evolution. Recalling that the ``elements of [a] primitive ontology are the stuff that things are made of'' \cite{allori}, i.e. the foundational objects of a theory which are assumed to have real status, the purpose of this Section is to explore the possibility of a primitive ontology for isolated physical phenomena, assuming that their evolution (the ``experience'' of Definition~\ref{def2}) is subordinate to the restricted information available to the phenomena themselves. In this respect, the notion of experience we propose is analogous to information theoretic approaches to the foundations of physics \cite{zell,hooft}.

\begin{mydef}
\label{def1}
An  {\bf isolated unit}  (or a unit) is a physical phenomenon that is affected by potential fields generated by other phenomena and that does not significantly affect its surrounding.
\end{mydef}
This definition of units gives a basic criterion to establish when a phenomenon counts as distinct, by stressing the asymmetry in its interactions with the surrounding.  The terms ``interaction'' and ``potential field'' are assumed to be given, primitive notions and in the following they will be clarified case by case in specific physical contexts. Note that, in principle, there is no restriction on the size or structure of a unit, as units are defined relationally. The expression ``does not significantly affect'' can be made precise by saying that the state of the surrounding in the presence of the unit cannot be distinguished from the state of the surrounding when the unit is absent. 

If we consider macroscopic changes of the surrounding, this definition can be seen as a variation on the view that changes in the ``position of things'' \cite[Chapter 19.2]{Bell} are the relevant variables of a meaningful physical theory. However, according to Definition~\ref{def1}, a unit is not isolated also when there are changes the microscopic state of another object (say, its spin value), and not only when there are macroscopic changes in the state of the surrounding.
\begin{assu}
\label{assu1}
Every unit  has a set of {\bf internal state variables} that define its physical properties and an associated {\bf internal time state} with respect to which changes of the internal state variables are defined. 
\end{assu}
Internal state variables include the list of physical properties that define the characteristics of a phenomenon needed to study its evolution. According to the interaction that is considered in the theory, different sets of internal state variables may be considered. For example, if we consider gravitational potential fields, we will take mass as one of the internal state variables of a unit, and if electromagnetic fields are considered, the charge of the unit will be an internal state variable, and its spin in more general settings. 

Changes in the values of the internal state variables (when such changes are permissible) are established with respect to an internal time state $\tau$. In this Section and in Section 3.1 it is helpful to think of $\tau$ as the proper time along the trajectory of the unit. We will see in Section 3.2 that $\tau$ needs to satisfy some specific mathematical constrains to be compatible with the framework of the units of experience, and we will give its analytical form, distinct from proper time, in Definition~\ref{tempogiusto} of Section 3.2.
 
 \bigskip
When looking at the evolution of the internal state variables of a unit, there is in general no way from their changes to reconstruct the surrounding in which they are evolving. For example, a particle with spin 
may change its spin value, which possibly implies the presence of a variable magnetic field, but the simple fact that the value has changed does not allow us to reconstruct the surrounding field.
However, a surrounding is implied nevertheless; note that two evolutions can be identical, and still need to be considered as distinct according to the environment in which they occur. We summarize this discussion in the definition of labelled experiences, which will become necessary when defining indistinguishable experiences and associated indistinguishable paths in Section 2.2.

\begin{mydef}
\label{def2}
A {\bf labelled experience} of a unit is the evolution of its internal state variables over some period of internal time, for a given surrounding environment, and a given spacetime localization (a path) within the surrounding.
\end{mydef}

For ease of reading we will usually drop the qualification ``labelled'' for experiences, and we will refer to them as ``experiences'' tout court.

\begin{mydef}
\label{def3}
A unit whose internal state variables cannot change with respect to internal time is a {\bf simple  unit}.
\end{mydef}
Isolated spinless elementary particles are the main examples of simple units, with internal state variables defined by mass and charge. Any unit that is not simple is a {\it structured unit}. Structured units include: isolated composite units, such as atoms, made of simple units; complex molecules and macroscopic, composite objects; isolated elementary particles that can have varying spin states; entangled elementary particles. While elementary particles are the main example of units we will work with, our theory is not restricted to microscopic objects, as long as the condition of isolation is satisfied. 

\begin{remark}
\label{generalUnits}
This paper will be concerned mostly with simple units, structured units will be addressed separately. Still, whenever a result will not be specifically restricted to simple units, we may assume it to be equally valid for structured units, with suitable modifications on the range and type of experiences. As already noted, in principle the primitive ontology we are building affirms the primacy of the notion of isolated units (whenever they arise) in understanding phenomena {\it irrespective of the size or complexity of the units}. Even though our main results in this paper concern spinless elementary particles, the framework of isolated units is not specifically a theory of elementary particles.
\end{remark}

\bigskip
\noindent
Note that a specific embodiment for internal time is provided by the hypothesis of De Broglie of an internal frequency being attached to any finite mass body \cite[Chapter 1]{deBroglie1}. In particular, de Broglie conjectured that for any body of finite rest mass $M_0>0$ there is a internal phenomenon (a ``clock'') of frequency $\nu_0=M_0c^2/h$ attached to the body itself, where $h$ is Planck's constant 
and $c$ is the speed of light. Alternatively, we can write $\omega_0=M_0c^2/\hbar$, if for simplicity we switch to units of measure such that angular frequency is measured in radians per second and with $\hbar=\frac{h}{2\pi}$ the reduced Planck constant. 

In Section 3.2 we will derive a relativistically invariant propagator of simple units that uses in an essential way the internal clock frequency and in Remark~\ref{alpha} we argue that the analytic form of the propagator requires that the internal clock frequency for the internal time state must be a multiple $\alpha$ of the one conjectured by de Broglie. In Section 4 we will then show that our theory reduces to standard quantum mechanics (under the conditions in which the latter applies) if we choose  $\alpha=\frac{1}{2}$.  Accordingly, already in this Section we redefine $\omega_0$ as follows: 
\begin{mydef}
\label{innerclock}
The {\bf internal clock frequency} $\omega_0$ of an isolated unit of rest mass $M_0>0$ is: 
\begin{equation}
\label{eq:alphaomega}
\omega_0=\frac{\alpha M_0c^2}{\hbar},
\end{equation}
where $\alpha$ is an appropriate scaling constant.
\end{mydef}

The physical reality of the internal clock frequency has been indirectly tested by electron channeling in silicon crystals \cite{testdeBroglie1,testdeBroglie2}. These experimental results seem consistent with a value of the internal frequency that is double the one predicted by de Broglie. This preliminary experimental discrepancy has no direct impact in what follows, and does not contradict our choice of $\alpha=\frac{1}{2}$ in Section 4,  as the internal clock frequency will always appear in our work in relation to internal time, and not time as measured in an external frame of reference.

Indeed, the  periodic representation of $\tau$ compatible with the internal clock frequency $\omega_0$ can be taken to be $\mathcal{P}(\tau)=e^{i\omega_0 \tau}$. Note that, in principle, any periodic function of period $\omega_0$ would be suitable to define $\mathcal{P}(\tau)$, in the framework we have described so far. The significance of the complex exponential in the setting of units of experiences will become clear when evaluating differences of internal time states for distinct trajectories of the unit, as we will do in the next section.

\begin{remark}
\label{debroglie}
The notion of internal clock frequency has been used in several works exploring the interfacing of quantum mechanics with the theory of relativity and in the presence of gravitational or more general potential fields  \cite{dolce1,dolce2,kastner2,mueller,refDeBroglie1}. The most comprehensive approach to quantum mechanics based on the de Broglie hypothesis can be found in the body of work exemplified by \cite{dolce1,dolce2}, where a Lorentz invariant description of elementary particles is given in terms of cyclic Minkowskian space-time coordinates associated to the internal clock frequency and in the context of the deterministic dynamics of one-dimensional classical closed strings vibrating in a four-dimensional spacetime. In particular, in \cite{dolce2} the classical evolution of all de BroglieÕs internal clock dynamics (that satisfy periodic boundary conditions in a cyclic time dimension) is proven to be equivalent to the Feynman path integral propagator in ordinary spacetime.
\end{remark}

As it is the case for all recent works that assume the de Broglie hypothesis, the crucial impact of the internal clock frequency in our framework is that it endows the notion of internal time state with a periodic representation. Nevertheless, the main objective of this work is to explore some of the concrete and most direct physical consequences of the primitive ontology of isolated units. Accordingly, the notion of isolated unit will continue to be the dominant thread in the way the internal clock frequency will be used, and in particular in establishing Lorentz invariance for the propagator of simple units in Section 3.2.

\subsection{Indistinguishability and Isolated Equivalence}

We established in Section 2.1 the basic notions of isolated unit and of experience of a unit. We now address how the type of internal state variables of a unit affect the possibility of internally distinguishing distinct labelled experiences. 

\begin{mydef}
\label{def4}
{\bf Indistinguishability.} 
Any two labelled experiences of a unit are internally indistinguishable if they cannot be distinguished through changes in the unit's internal state variables over internal time. The corresponding spacetime localizations of the two experiences are defined as the internally {\bf indistinguishable paths} of the unit.
\end{mydef}
With Definition~\ref{def4} we move to a crucial point: when the internal state variables are not changed (or are equally changed) by any pair of experiences, there is no way for the unit to establish in isolation which of the two has happened, they are not distinguishable. 

Since the unit is isolated and it is the only phenomenon validating the reality of a specific evolution of its internal states, indistinguishable experiences that cannot be physically discriminated before interaction are all valid, real experiences for the unit itself. For a simple unit, whose internal state variables are unchanging, all externally labelled and distinct experiences will be internally indistinguishable and therefore physically valid in isolation.

Specifically, in the experience of an isolated unit,  {\it its motion at any instant can be taken to be in any direction and at any velocity}. A simple unit has no ability to distinguish these distinct paths, since its internal state variables are unchanging over internal time. Therefore, all these possibilities of motion are fully consistent with its experience, even in the presence of potential fields that exert a force on the unit, as long as the unit itself does not need to match its experience with other units. Because motion at each instant could be at any velocity and in any direction, a simple unit experiences motion on any continuous, but not differentiable path. Even moving at a speed higher than light is logically allowed in this setting, as long as the unit is not forced to confirm this possibility with the external environment\footnote{When dealing with structured units, care must be given to the identification of the appropriate set of indistinguishable paths that change the internal state variables in the same way. This set will be generally smaller than the corresponding set for simple units.}.  

\begin{remark}
Note that the ``reality of possibility'' is a basic tenet of the transactional interpretation of quantum mechanics \cite{cramer,kastner}, where it is derived from the formalism of quantum mechanics itself. In the setting of the units of experience, the reality of the entire range of indistinguishable possibilities of motion is derived from the notion of isolation and from the type of internal state variables of the unit, before establishing a formal connection with quantum mechanics. In Section 3.1 we will show that the impact of the whole range of indistinguishable paths of a unit in interaction is always mediated by the assessment of the significance of each path with respect to all the others. In this respect, indistinguishable paths are not to be considered classical paths, because we cannot fully separate them from each other in determining their relational impact on external events such as localization at a specific spacetime point. 

We also note that the notion of indistinguishability, though set here as a general property of the restricted experience of isolated units, has a direct antecedent in the general sum-over-histories formalism of  quantum theories with its distinction of variables in observables and unobserved labels (see for example \cite{hartle}). In this formalism, probability amplitudes in path integrals are summed over the range of unobserved labels of a given phenomenon, if a corresponding ``experiment does not determine the end of a history precisely, as most will not'' \cite{hartle}. Effectively, unobserved labels can be considered in our terminology as states defining indistinguishable labelled experiences.
\end{remark}

This paper does not specifically address the mechanism of interaction, still, our framework would not be complete without the inclusion of a general prescription on how the labelled experiences of units, described by the evolution of their internal state variables, are affected in interaction. To this purpose, we define a notion of matching of experiences.

\begin{assu}
\label{assu4}
{\bf Matching of experiences.} When units interact, they match their labelled experiences to be reciprocally compatible.
\end{assu}
This assumption basically asserts that the range of indistinguishable paths of a unit is reset every time they interact. Consider, for example, the range of all indistinguishable paths of an isolated simple unit $S$; we have seen in the discussion of Definition~\ref{def4} that all types of motions are allowed for such unit in isolation. Assume now that a macroscopic structured unit $U_M$ (say, a measuring device) does not have multiple indistinguishable paths.  Then any interaction of $S$ with $U_M$ will reset the range of labelled experiences of $S$, localizing them to the position of their interaction with respect to $U_M$, and only indistinguishable paths of $S$ compatible with this localization will be preserved. Without this assumption, it would not be meaningful to speak of the propagation of a unit from a specific point $A$ to another point $B$, as we will do in Section 4.

The definition of matching of experiences has also important implications on the isolated status of units. If a unit $U_1$, because of some changes in its internal state variables, experiences being no longer separated from unit $U_2$, then unit $U_2$ will be no longer isolated, even if in its experience there is no change in its internal state variables. We will  refer to this observation in Remark~\ref{isolation1} and Remark~\ref{radiation} when evaluating under which conditions moving charged particles can be considered isolated units.

By assuring that every time units interact their labelled experiences are reciprocally consistent, we preserve a form of objectivity,  while at the same time acknowledging that such objectivity is always mediated by the act of matching the experiences of the units to each other. Note that the moment two distinct units interact, i.e. they exert a reciprocal influence on each other,  they cease to be distinct in the sense of Definition 1 for the whole duration of their interaction. Note also that no limitation is put on the conservation of the number of units before and after the interaction.

\bigskip
\noindent
The framework of units of experience allows a reformulation and an extension of the equivalence principle, that  emphasizes its relation to indistinguishability. We introduce therefore the following principle:
\begin{prin}
\label{prin2}
{\bf Isolated equivalence:} If a unit cannot internally distinguish the labelled experience of being in a stationary frame in a potential field  from the labelled experience of being at rest in a non-inertial frame with uniform acceleration, then the two experiences are physically equivalent.
\end{prin}

The theory of general relativity was motivated by the realization that locally there is no way to distinguish a frame of reference in uniformly accelerated motion from one that is stationary within a corresponding gravitational field. However, this is true only in the limit of infinitesimal systems, that do not experience the discriminating tidal effects in a gravitational field \cite{Ohanian}.  Under these limiting conditions, a single isolated charged particle can be effectively considered a simple unit (notwithstanding possible spin states). For such a unit, a frame in uniformly accelerated motion will also be indistinguishable, for example, from a static frame in a Coulomb field. The unit will not be able to discriminate the nature of the field by detecting changes to its internal state variables.  

Isolated equivalence assumes that  the standard equivalence principle can be extended to all cases, such as this, where the unit cannot discriminate the nature of the field acting on it\footnote{We focus on simple units in the justification of the principle of isolated equivalence, because of their significance in subsequent Sections. However, the principle is  formulated to apply to general isolated units.}, {\it if only in the isolated experience of a simple unit}.

Indeed, in the case of simple units,  to be ``physically equivalent'' refers essentially to dilation effects on internal time, since all other internal state variables are unchanging; internal time dilation is the same for the two indistinguishable experiences with respect to time in an external frame of reference. 
Note that a surrounding is always implied when dealing with labelled experiences, in that we still need an external frame of reference to be able to quantify time dilation\footnote{An external frame of reference can be defined with respect to the position and state of motion of the unit itself at the time of matching of experiences, as we explain in Section 3.2.}. The implicit role of an external frame of reference makes sure that only non-inertial frames with the same acceleration as the one induced on the unit by the potential field will be properly defined and physically equivalent, despite the fact that all non-inertial frames of reference are indistinguishable for a simple unit.

\begin{remark}
\label{isolation1}
Isolated equivalence does not contradict the fact that static reference frames in a general potential field are not in general equivalent to uniformly accelerated frames. Such equivalence is only valid for a simple isolated unit of experience. The motion of continuously interacting units in an electromagnetic field will depend as expected on relativistic effects on the mass\footnote{Similarly, the Abraham-Lorentz-Dirac force (due to the interaction of the charged unit with its own electromagnetic field and to the corresponding radiating field) does not break the equivalence in the isolated perspective since it cannot affect the internal state variables of  a simple unit. Such self-force can be interpreted in the experience of the unit as an effect of a varying potential field.}. In particular,  radiation may affect the unit by making it no longer isolated, as we argue in Section 5. Lorentz invariance of electromagnetic laws is preserved in this case, consistently with the standard equivalence principle and its experimental validations \cite{centennial}.
\end{remark}

Unlike the standard equivalence principle of general relativity, isolated equivalence is a conditional principle that depends on the inability of the unit to distinguish experiences. The dependence of isolated equivalence on the type of internal state variables of the unit, and especially on the unit being isolated, sets our contribution apart from other generalizations of the equivalence principle to general fields such as the one developed in \cite{friedman}. We will show in Section 4 that this dependence is also the key for the application of isolated equivalence to quantum mechanics. 

\section{Propagation of Simple Units}

\subsection{Relative Significance and Time Invertibility}

We now explore the question of {\it propagation } of a simple unit $S$ in a reference frame, that is, of how a unit $S$ that is localized at a point $A=(x_a,t_a)$ is then found at  point $B=(x_b,t_b)$. The localization of the simple unit at these two points is established by assuming interaction by matching of experiences with other units $U_A$ at $A$ and $U_B$ at $B$. Even though this question applies to any isolated unit, we assume here that we deal only with simple isolated units such as spinless elementary particles, with mass and charge as internal state variables.

\begin{remark}
\label{old3_1}
As already noted in Section 2.1, we  continue in this Section 3.1 to assume that $\tau$ is analogous to proper time and we use the internal clock frequency $\omega_0$ defined from rest mass for the periodic representation $\mathcal{P}(\tau)$. In Section 3.2  we give a specific representation for the internal time state, distinct from proper time, and we define a relativistic form for the internal clock frequency.
\end{remark}

To take all indistinguishable experiences and their corresponding paths as real means that, in interacting with another unit, they all should have a physical impact. By considering the relation of indistinguishability and isolated equivalence of the experiences of a unit, we clarify the modalities of this impact in Definitions ~\ref{def6} and ~\ref{def7} that together establish the notion of relative significance of the experience of a unit. They quantify to which extent a given experience of an isolated unit can be trusted to be physically significant in the process of matching of experiences.

\begin{mydef}
\label{def6}
Let $\gamma_1$ and $\gamma_2$ be two indistinguishable paths of a (simple) unit $S$ starting at $A$ and ending at $B$. Assume $S$ takes internal time $\tau_1$ to reach $B$ from $A$ along $\gamma_1$,  and $\tau_2$ along $\gamma_2$. The {\bf measure of equivalence} $\mathcal{M}(\gamma_1,\gamma_2)$ of the labelled indistinguishable experiences associated to the two paths $\gamma_1$ and $\gamma_2$ is the periodic representation of the difference of their internal time durations:
\begin{equation}
\label{eq:timediff}
\mathcal{M}(\gamma_1,\gamma_2)=\mathcal{P}(\tau_1-\tau_2)=e^{i\omega_0 (\tau_1-\tau_2)}.
\end{equation}
\end{mydef}
Since for indistinguishable experiences internal state variables evolve in the same way, internal time differences become the only way to partially assess the impact of the surrounding on distinct indistinguishable experiences of the unit. With respect to the experience of the unit at $B$, any time interval, including the time difference $\tau_1-\tau_2$,  can only be assessed through the periodization of internal time, as  $\mathcal{P}(\tau_1-\tau_2)=e^{i\omega_0 (\tau_1-\tau_2)}$. This representation of time differences assures that they are compared, across indistinguishable paths, with periodicity $\omega_0$ corresponding to the unit's internal clock frequency.

The measure of equivalence assesses the differential impact of the surrounding on the indistinguishable experiences of the unit. This impact may be due to the geometric characteristics of each path (say, its length), and to time dilation effects due to potential fields via isolated equivalence. 

\begin{remark}
As anticipated in Section 1.1 when introducing the periodic representation of internal time $\mathcal{P}(\tau)$, the relevance of the complex exponential function for units of experience is in great part due to the possibility of expressing the periodic representation of the time difference $\tau_1-\tau_2$ in terms of the periodic representation of  $\tau_1$ and $\tau_2$ respectively, via the relation $e^{i\omega_0 (\tau_1-\tau_2)}=e^{i\omega_0 \tau_1}e^{-i\omega_0 \tau_2}$. This property is the key in Equation~\eqref{eq:splitint} to an analytical form for the relative significance of the unit going from $A$ to $B$. 
\end{remark}

We are now ready to quantify the significance of each experience in interactions.
\begin{mydef}
\label{def7}
The {\bf relative significance} of an experience $E$ with respect to a set $\mathcal S$ of experiences indistinguishable from $E$ is the sum of all pairwise measures of equivalence of $E$ with the elements of $\mathcal S$.
\end{mydef}
The extent to which a specific experience of a unit can be trusted to be physically meaningful in an interaction is validated by all other indistinguishable experiences that are equally involved in the interaction. Therefore the totality of all internal time differences of an experience with its set of indistinguishable experiences becomes a proxy for the significance of the experience itself. This concept will play a crucial role in understanding the propagation of units in Section 3.

Note that a notion of ``distinctiveness'' of the views of elements of a system plays an important role in \cite{smolin}, similar to the one played in our framework by the relative significance and the measure of equivalence. 

\bigskip
\noindent
Consider now several indistinguishable paths $\{\gamma_1,\dots,\gamma_n\}$ starting at $A$ and ending at $B$ with internal time durations $\{\tau_1,\dots,\tau_n\}$ respectively. In Definition~\ref{def7} the relative significance of an indistinguishable experience $E$ of the unit $S$ is defined as the sum of all pairwise measures of equivalence of experience $E$ with respect to the other indistinguishable experiences. Each path $\gamma_i$, $i=1,...,n$ identifies one experience of $S$, therefore the (normalized) relative significance of  $\gamma_i$ with respect to the remaining indistinguishable paths is:
\begin{equation}
\label{eq:Wgamma}
W(\gamma_i)=\frac{1}{n-1}\sum_{j\neq i} \mathcal{M}(\gamma_i,\gamma_j)=\frac{1}{n-1}\sum_{j\neq i} e^{i\omega_0(\tau_i-\tau_j)}.
\end{equation}
We can now express the normalized relative significance $W(A,B)$ of the unit going from $A$ to $B$ as the sum of the relative significance of all paths starting at $A$ and ending at $B$, scaled by n:
\begin{equation}
\label{eq:W_AB}
W(A, B)=\frac{1}{n}\sum_i  W(\gamma_i).
\end{equation}

We can calculate the relative significance of a path $\tilde \gamma_{AB}$ in the limit of infinitely many paths in analogy to Equation~\eqref{eq:Wgamma}, by assuming a suitable measure $D\gamma_{AB}$ on the paths themselves. We define $\tau_{\gamma_{AB}}$ to be the total internal time along a generic path $\gamma_{AB}$ and we write:
\begin{equation}
W(\tilde \gamma_{AB})=\int e^{i\omega_0 (\tau _{\tilde \gamma_{AB}}-\tau_{\gamma_{AB}})} D\gamma_{AB}.
\end{equation}
The relative significance of the unit going from $A$ to $B$ becomes then the integral over $\tilde \gamma_{AB}$ of $W(\tilde \gamma_{AB})$:
\begin{equation}
\label{eq:splitint}
\begin{split}
W(A, B)=\int (\int e^{i\omega_0 (\tau _{\tilde \gamma_{AB}}-\tau_{\gamma_{AB}})} D\gamma_{AB}) D\tilde \gamma_{AB} \\=
\int e^{i\omega_0 \tau_{\tilde \gamma_{AB}}} D\tilde \gamma_{AB} \cdot \int e^{-i\omega_0 \tau_{\gamma_{AB}}} D\gamma_{AB},
\end{split}
\end{equation}
where we assume that the integrals can be separated. We then set $\int_{\gamma_{AB}}d\tau$ to be an integral parametrization of the total internal time $\tau_{\gamma_{AB}}$ along the path $\gamma_{AB}$ starting at $A$ and ending at $B$ and we write:

\begin{equation}
\begin{split}
\label{eq:doubleint}
W(A, B)=
\int e^{i\omega_0 \int_{\tilde \gamma_{AB}}d\tau}D\tilde \gamma_{AB} \cdot
\int e^{-i\omega_0 \int_{\gamma_{AB}}d\tau} D\gamma_{AB}.
\end{split}
\end{equation}

Now, in general the two integrals in Equation~\eqref{eq:doubleint} will be different. We note that
\begin{equation}
\begin{split}
\label{eq:timesymmetry}
\int e^{-i\omega_0 \int_{\gamma_{AB}}d\tau} D\gamma_{AB}=\int e^{i\omega_0 \int_{\gamma_{BA}}d\tau} D\gamma_{BA}
\end{split}
\end{equation}
which can be interpreted to mean that $W(A,B)$ depends both on paths from $A$ to $B$ and on those from $B$ to $A$. 

The time-symmetric structure of  $W(A,B)$ is reminiscent of the dependence of a quantum system on the boundary conditions in the future and in the past as posited by  the two-state-vector formalism of quantum mechanics \cite{aha1,aha2}, or by the transactional interpretation of quantum mechanics\cite{kastner}.

We expect that the expression in Equation~\eqref{eq:timesymmetry} will be significant when exploring causality constraints for structured units. However, in this paper we make the following simplifying assumption that leads to a mathematical form for $W(A,B)$ closer to the  standard quantum mechanical propagator.

\begin{condition}
\label{timeInv}
{\bf Time invertibility.} Given the space $\Gamma_{AB}$ of indistinguishable paths from $A$ to $B$, there is a map 
$\Sigma: \Gamma_{AB}\rightarrow \Gamma_{AB}$ such that: 
\begin{enumerate}
\item the image of $\Sigma$ is dense in $\Gamma_{AB}$; 
\item the measure $D\gamma_{AB}$ in 
\begin{equation}
\int e^{-i\omega_0 \int_{\gamma_{AB}}d\tau} D\gamma_{AB}
\end{equation}
is invariant (up to sign) with respect to $\Sigma$; 
\item and such that 
\begin{equation}
\label{eq:flip_int}
\int_{\Sigma(\gamma_{AB})}d\tau=-\int_{\gamma_{AB}}d\tau. 
\end{equation}
\end{enumerate}
\end{condition}
Under Condition~\ref{timeInv} we have:
\begin{equation}
\begin{split}
 \label{eq:same_int}
 \small
W(A, B)=\int e^{i\omega_0 \int_{\gamma_{AB}}d\tau}D\gamma_{AB} \cdot
\int e^{i\omega_0 \int_{\Sigma(\gamma_{AB})}d\tau} D\Sigma(\gamma_{AB})=\\
\pm (\int e^{i\omega_0 \int_{ \gamma_{AB}}d\tau}D\gamma_{AB})^2.
\end{split}
\end{equation}

If we want to compare the relative significance of different sets of indistinguishable paths, as for example the set of paths from $A$ to $B$ with those from $A$ to $B'$, we can provide a partial ordering of relative significance by taking the module of $W$ in Equation~\eqref{eq:same_int}:
\begin{equation}
\label{eq:W_ordered}
|W(A, B)|=|\int e^{i\omega_0 \int_{\gamma_{AB}}d\tau} D\gamma_{AB}|^2
\end{equation}
This partial ordering of the relative significance makes sense only from an external perspective. Indistinguishable paths cannot be separated into subgroups unless distinct matchings of experiences are considered.

If we scale uniformly $|W(A, B)|$ to make it into a probability distribution, we can equate it with the likelihood $P(A,B)$ of a simple unit that interacted at $A$ with another unit to then interact again with another unit at $B$. Up to a rescaling hidden in the measure of the integral, we write:
\begin{equation}
\label{eq:P}
P(A,B)=|\int e^{i\omega_0 \int_{\gamma_{AB}}d\tau} D\gamma_{AB}|^2
\end{equation}

As much as the formal connection of $|W(A, B)|$ with the computation of probabilities via path integrals in quantum mechanics is obvious, conceptually the process that led to
Equation~\eqref{eq:P} is very different. 

We established how localization of a simple unit at two spacetime points $A$ and $B$ affects the relative significance of all compatible indistinguishable paths. We defined the relative significance of going from $A$ to $B$ and finally we imposed an ordering on the relative significance of distinct choices of the pair of points $(A,B)$. The interplay of indistinguishability and equivalence naturally led to a representation of the likelihood $P(A,B)$ of propagating from $A$ to $B$ as the square of the module of a complex number.

\begin{remark} 
\label{old3_2}
The description of simple units by their internal time and its phase representation is reminiscent of the description of light propagation suggested in \cite{Feynman1}, with its stress on stopwatches associated to photons paths. However, the dependence of probability amplitudes from elapsed time in the photon detection is not a fundamental characteristic of the physics of photons. It is rather a consequence of the excitation and decaying processes of atoms involved in emitting and detecting the photons themselves \cite[Section~4]{Field1}, \cite{Field2}. We also note that the framework of simple units does not apply to photons in its current form, but rather to elementary particles of nonzero rest mass.
\end{remark}

\subsection{A Relativistically Invariant Propagator for Simple Units.} 

As noted in Section 2.2,  a simple isolated unit $S$ experiences motion on the set of all its indistinguishable paths. Accordingly, it is not possible to reduce the integral in Equation~\eqref{eq:P} only to paths that are time-like: the velocity $v$ measured along the path must be allowed in principle to be higher than the speed of light $c$.  While in general allowing paths with speed greater than the speed of light  leads to acausal effects \cite{redmount}, causality is definable only when units interact, and instead here we allow acausal paths only for simple isolated units.The physical mechanism that enforces $v<c$ for a simple unit must be relational, and dependent on the matching of experiences. 

\begin{remark}
On this last point, we note that a relational interpretation of quantum mechanics (RQM) was first introduced in \cite{rovelli}. RQM assumes a radical form of relationality according to which all physical quantities continue to have different values in the perspective of different reference systems. In our setting, the experiences of the units directly involved in the matching are privileged and determine which outcome is mutually compatible. Relationality ultimately is dependent on the type of internal state variables of the specific units that are matching their experiences.
\end{remark} 
Regardless, to be able to apply properly the principle of isolated equivalence, it is still necessary to consider possible relativistic effects on the scaled module of the measure of relative significance $P(A,B)$, that we equated in Equation~\eqref{eq:P} to the likelihood of a unit propagating from $A$ to $B$. 

The consideration of relativistic effects will require us to choose a specific form for the internal time differential $d\tau$ in Equation~\eqref{eq:P}, and in turn of internal time intervals $\tau_{\gamma_{AB}}= \int_{\gamma_{AB}}d\tau$, such that: it respects the properties of simple isolated units; it satisfies the time invertibility Condition~\ref{timeInv}; and it ensures relativistic invariance for $P(A,B)$ for the frames of reference that are physically meaningful for the unit $S$ under consideration.

\paragraph{Proper time does not satisfy Condition~\ref{timeInv}.} A possible choice for $d\tau$ in the exponent of Equation~\eqref{eq:P} could be to set it equal to the square root of the Lorentzian line element, corresponding to relativistic invariant proper time, i.e. $d\tau=(ds^2/c^2)^{1/2}$.  This is the choice we implicitly made throughout Section 2 and 3.1 and it is standard in Lagrangian approaches to relativity \cite[Chapter~3.19]{GE1}, \cite{rindler}.

However, $d\tau=(ds^2/c^2)^{1/2}$ is not truly compatible with the framework of units of experience, and a different expression for $d\tau$ and $\tau_{\gamma_{AB}}$ will be given in Definition~\ref{tempogiusto}.

The first issue with the choice $d\tau=(ds^2/c^2)^{1/2}$ is that it is explicitly incompatible with Condition~\ref{timeInv}. If $d\tau=(ds^2/c^2)^{1/2}$, then for all paths  $\tau_{\gamma_{AB}}=\int_{\gamma_{AB}}d\tau=a+ib$, with $a,b>0$ (if in addition a path has some portions with $v>c$ then $b\neq0$). Under these conditions, we cannot find a map of the space of paths onto itself to account for the negative sign in the integral on the right hand side of  Equation~\eqref{eq:doubleint}. 

Second, this choice of $d\tau$ is not fully compatible with the setting of units of experience. To see this, recall that by definition indistinguishable paths cannot be discriminated in the experience of the unit. Suppose now we have a path $\gamma$ with internal time interval $\tau_{\gamma_{AB}}=a+ib$, then the periodized internal time would be $e^{i\omega_0 (a+ib)}=e^{-\omega_0 b}e^{i\omega_0 a}$. The exponentially decaying factor $e^{-\omega_0 b}$, would break the uniformity of internal time states across indistinguishable paths: simply by having $e^{-\omega_0 b}<1$ we would be able to ascertain whether that path had in its past portions with $v>c$. 

Note that this second issue with $d\tau=(ds^2/c^2)^{1/2}$ stands also when the relative significance $W(A,B)$ is defined according to its form in Equation~\eqref{eq:doubleint}, that does not depend on the time invertibility Condition~\ref{timeInv}.

We conclude that, despite its relativistic invariance, proper time $d\tau=(ds^2/c^2)^{1/2}$ cannot be used to define internal time: it is not compatible with the time invertibility Condition~\ref{timeInv} on the analytic form of the relative significance; and it does not  guarantee for simple units the indistinguishability of paths, because of the discriminant introduced by complex valued internal time values dependent on the path. 

\begin{remark}
\label{old3_3}
Note that relativistic path integrals based on $d\tau=(ds^2/c^2)^{1/2}$ recover causality for distances that are large enough \cite{redmount}, exactly because of the exponential decay $e^{-b}$ associated to paths with $v>c$. Nevertheless the breakdown of causality at short distances is unavoidable even for this choice of $d\tau$. A similar conclusion was already reached in \cite{hartle}, where it is shown that simple quantum systems with no preferred time parameter lead to path integral propagators that must include acausal paths. 

\end{remark}

\paragraph{Frames of reference at points of matching of experiences.} To define a suitable $d\tau$ that preserves relativistic invariance for $P(A,B)$, we need first to establish which frames of reference are physically meaningful for interacting units of experiences. 

Recall that the simple unit $S$ under consideration is localized at $A$ and $B$ by interaction respectively with units $U_A$ and $U_B$. 
We argue that only from the perspective of the experience of units such as $U_A$ and $U_B$\footnote{Or units that are in an unbroken chain of matches of experience with $U_A$ and $U_B$.} it is possible to define an external frame of reference that is relevant to the propagation of $S$ and whose relative velocity with respect to $S$ is {\it uniquely} defined.

Indeed, $S$ is localized at $A$ and all its indistinguishable paths will agree on the position and velocity of $S$ with respect to $U_A$.
Denote by $u$ the relative speed of $U_A$ and $S$ at $A$, we can then uniquely define, with respect to $S$, the frame of reference $\mathcal{F}_u$ at rest with $U_A$ at $A$ (the same argument can be made for frames of references defined by $U_B$ at $B$).

We call this privileged set of frames of references defined at matching points the {\it matching frames of reference} of the unit $S$, and we only refer to such frames in the following arguments.\footnote{We simplify our analysis in this Section by assuming that the interaction at $B$ is with a unit $U_B$ that is also at rest in $\mathcal{F}_u$.} We can now define:

\begin{mydef}
\label{properDeB}
The {\bf relativistic internal clock frequency} of a unit $S$ as observed from a matching frame of reference $\mathcal{F}_u$ in motion at speed $u$ with respect to unit $S$ as localized at $A$ is:
\begin{equation}
\label{eq:omegaR}
\omega_u=(1-\frac{u^2}{c^2})^{-1/2}\omega_0
\end{equation}
\end{mydef}
All experiences of the unit are defined with respect to the periodic representation of internal time $\mathcal{P}(\tau)=e^{i\omega_0 \tau}$, dependent on the rest mass $M_0$ of a unit. It follows that relativistic effects on $\omega_0$ must be defined directly in terms of relativistic effects on the rest mass $M_0$.
\begin{remark}
\label{old3_4}
Note the discrepancy of our definition with respect to the original suggestion of internal clock of de Broglie \cite[Chapter 1.1]{deBroglie1}. De Broglie, building on  the inverse relation of period and frequency and on time dilation, postulates that 
$$\omega_u=(1-\frac{u^2}{c^2})^{1/2}\omega_0$$
while the relativistic form in Equation~\eqref{eq:omegaR} is reserved in de Broglie's work to the  {\it matter wave} associated to the body and propagating in space. This suggestion of de Broglie does not allow the construction of a relativistic propagator for simple units (see also Remark \ref{old3_5}).
\end{remark}

We now define a particular choice of internal time differential $d\tau$ compatible with the setting of units of experience, with the corresponding likelihood $P(A,B)$ and internal time path integral.

We denote by $s$ spacetime point coordinates, and we parametrize $s$ on all indistinguishable paths with respect to time $t'$ as measured in a matching frame of reference $\mathcal{F}_u$. We further denote by $\frac{ds}{dt'}*\frac{ds}{dt'}$ the Lorentzian inner product of $\frac{ds}{dt'}$ with itself.

\begin{mydef}
\label{tempogiusto}
Given a matching frame of reference $\mathcal{F}_u$ in motion at speed $u$ with respect to the simple unit $S$ as localized at $A$, the internal time differential  is defined as:
\begin{equation}
\label{eq:int_time_diff}
d\tau=\frac{1}{c^2}\frac{ds}{dt'}*\frac{ds}{dt'}dt'.
\end{equation}
The corresponding internal time interval $\tau_{\gamma_{AB}}$ along the path $\gamma_{AB}$ is: 
\begin{equation}
\label{eq:int_time}
\tau_{ \gamma_{AB}}=\int_{\gamma_{AB}}\frac{1}{c^2}\frac{ds}{dt'}*\frac{ds}{dt'}dt'.
\end{equation}
\end{mydef}

\begin{remark}
\label{old3_5}
Recall that with respect to time $t'$ as measured in an external matching frame of reference we can write 
$$(ds^2/c^2)^{1/2}= \frac{(ds^2/c^2)^{1/2}}{dt'}dt'=(\frac{1}{c^2}\frac{ds}{dt'}*\frac{ds}{dt'})^{1/2}dt'.$$
The internal time differential in Definition~\ref{tempogiusto} is simply a frame dependent representation of the differential of proper time where the square root in $(\frac{1}{c^2}\frac{ds}{dt'}*\frac{ds}{dt'})^{1/2}$ has been removed. We will show in Proposition 1 that, even though the internal time differential is frame dependent, 
the corresponding periodized internal time intervals $\mathcal{P}(\tau_{\gamma_{AB}})$ are relativistically invariant for matching frames of reference. In turn, $\mathcal{P}(\tau_{\gamma_{AB}})$ is the only physical quantity that is significant for computing the measurable probability distribution $P(A,B)$ we give in  Equation~\eqref{eq:finalP}. Note that the original suggestion of de Broglie on the Lorentzian transformation of the internal clock frequency (reported in Remark \ref{old3_4}) would not  allow relativistic invariance for $\mathcal{P}(\tau_{\gamma_{AB}})$. 
\end{remark}
Definition~\ref{tempogiusto} allows us to further define the corresponding expressions for the propagator of a simple unit $S$.
\begin{mydef}
\label{squarepath}
The likelihood $P(A,B)$ of $S$ propagating from $A$ to $B$  is defined in $\mathcal{F}_u$ as\footnote{This definition follows closely the one of the scaled relative significance in Equation~\eqref{eq:P}.}:
\begin{equation}
\label{eq:finalP}
P(A,B)=|\int_{ \gamma_{AB}} e^{i\omega_u \int_{\gamma_{AB}} \frac{1}{c^2}\frac{ds}{dt'}*\frac{ds}{dt'}dt'} D\gamma_{AB}|^2.
\end{equation}
The {\bf internal time path integral} (or {\bf the propagator}) associated to $P(A,B)$ is:
\begin{equation}
\label{eq:innerInt}
I(A,B)=\int_{ \gamma_{AB}} e^{i\omega_u \int_{\gamma_{AB}}\frac{1}{c^2}\frac{ds}{dt'}*\frac{ds}{dt'}dt'} D\gamma_{AB}.
\end{equation}
\end{mydef}
\paragraph{A relativistic periodic representation of internal time.} One of the main conclusions of this Section is the following proposition:
\begin{proposition}
Let the periodic representation of the internal time interval $\tau_{\gamma_{AB}}$ of a single unit along a path $\gamma_{AB}$ be:
$$\mathcal{P}(\tau_{\gamma_{AB}})=e^{i\omega_u \int_{\gamma_{AB}}\frac{1}{c^2}\frac{ds}{dt'}*\frac{ds}{dt'}dt'}$$
then $\mathcal{P}(\tau_{\gamma_{AB}})$ and the integrals $P(A,B)$ and $I(A,B)$, as in  Definition~\ref{squarepath}, are all relativistically invariant with respect to matching frames of reference defined at the points of interaction. 
\end{proposition}
\paragraph{Proof.} It is enough to show relativistic invariance of the argument of the exponent in $\mathcal{P}(\tau_{\gamma_{AB}})$, namely the quantity: 
\begin{equation}
\label{eq:T}
\mathcal{T}_u(\gamma_{AB})=\omega_u \int_{\gamma_{AB}}d\tau=\omega_u \int_{\gamma_{AB}}\frac{1}{c^2}\frac{ds}{dt'}*\frac{ds}{dt'}dt'.
\end{equation}
We denoted by $t'$ a parametrization of time in $\mathcal{F}_u$, and similarly we denote by $t$ a parametrization of time in  $\mathcal{F}_0$, the frame of reference at rest with respect to $S$ at $A$. From the perspective of $\mathcal{F}_0$, $t'=\epsilon t$ with $\epsilon=(1-\frac{u^2}{c^2})^{-1/2}$,  and $\omega_u=\epsilon\omega_0$,
therefore:
\begin{equation}
\begin{split}
\label{eq:calc_relative}
\mathcal{T}_u(\gamma_{AB})=\omega_u \int_{\gamma_{AB}}\frac{1}{c^2}\frac{ds}{dt'}*\frac{ds}{dt'}dt'=\\
\epsilon \omega_0 \int_{\gamma_{AB}} \frac{1}{c^2}(\frac{ds}{dt}\frac{dt}{dt'}*\frac{ds}{dt}\frac{dt}{dt'})\frac{dt'}{dt}dt=\\
\epsilon \omega_0 \int_{\gamma_{AB}} \frac{1}{c^2}(\frac{ds}{dt}\epsilon^{-1}*\frac{ds}{dt}\epsilon^{-1})\epsilon dt=\\
\epsilon \omega_0 \int_{\gamma_{AB}} \frac{1}{c^2}\frac{ds}{dt}*\frac{ds}{dt}\epsilon^{-1} dt=\\
\omega_0  \int_{\gamma_{AB}} \frac{1}{c^2}\frac{ds}{dt}*\frac{ds}{dt}dt=
\mathcal{T}_0(\gamma_{AB}).
\end{split}
\end{equation}

For any speed $u$, $\mathcal{T}_u(\gamma_{AB})$ is identical with $\mathcal{T}_0(\gamma_{AB})$, i.e. the same quantity as defined in the matching frame of reference $\mathcal{F}_0$ at rest with respect to unit $S$ at $A$, and therefore  $\mathcal{T}_u(\gamma_{AB})$ and  $\mathcal{P}(\tau_{\gamma_{AB}})$ are relativistically invariant. Since $P(A,B)$ and $I(A,B)$ are also defined in terms of $\mathcal{T}_u(\gamma_{AB})$, they are relativistically invariant as well. 
\qed

\bigskip
We will show now that the choice of $\tau_{\gamma_{AB}}$ in Definition~\ref{tempogiusto} is compatible with the experience of simple units, and also with the time invertibility Condition~\ref{timeInv}. We work for simplicity in the frame of reference $\mathcal{F}_0$.

First, note that it may happen that the internal time interval $\tau_{\gamma_{AB}}=\int_{\gamma_{AB}} \frac{1}{c^2}\frac{ds}{dt}*\frac{ds}{dt}dt<0$, but the periodization $\mathcal{P}(\tau_{\gamma_{AB}})=e^{i\omega_0\tau_{\gamma_{AB}}}$ makes the distinction of positive and negative values of $\tau_{\gamma_{AB}}$ meaningless, so that no inconsistencies in the experience of the unit arise. Indeed, in the definition of indistinguishability, we considered changes in the internal state variables with respect to internal time, irrespective of the direction of flow of the internal time state.

We now address the compatibility of $\tau_{\gamma_{AB}}=\int_{\gamma_{AB}} \frac{1}{c^2}\frac{ds}{dt}*\frac{ds}{dt}dt$ with the time invertibility Condition~\ref{timeInv}. 

\begin{proposition}
For every path $\gamma_{AB}$ of total internal time $\tau$, there is a path $\gamma'_{AB}$ of total internal time $-\tau$ as close as we want to $\gamma_{AB}$ itself, such that the map $\Sigma(\gamma_{AB})=\gamma'_{AB}$ satisfies the time invertibility Condition~\ref{timeInv}.
\end{proposition}
A full proof of this Proposition is beyond the scope of this paper, as the technical nature of such proof  contrasts with the informal way we have handled path integrals so far. However, we sketch here a justification. 
\paragraph{Sketch of Proof.} Let us work in the frame of reference $\mathcal{F}_0$. To show that $\tau_{\gamma_{AB}}=\int_{\gamma_{AB}} \frac{1}{c^2}\frac{ds}{dt}*\frac{ds}{dt}dt$ satisfies Condition~\ref{timeInv}, consider a piecewise linear path $\gamma_{(AB,N)}$ made of $N$ small segments $\lbrack A_i,A_{i+1}\rbrack$ with $A_1=A$ and $A_N=B$ and such that all segment end points are on $\gamma_{AB}$. 

Let $\tau_{\lbrack A_i,A_{i+1}\rbrack}=\delta \tau_i>0$. For small $\delta \tau_i$ we can approximate $\delta \tau_i=\frac{1}{c^2}\frac{\delta s_i}{\delta t_i}*\frac{\delta s_i}{\delta t_i}\delta t_i=\frac{1}{c^2}\frac{\delta s_i^2}{\delta t_i^2}\delta t_i$. Assuming without loss of generality that $\delta t_i>0$, we conclude that $\delta s_i^2>0$, i.e. $\lbrack A_i,A_{i+1}\rbrack$ is a time-like interval.

We can now build two connected segments $\lbrack A_i,H_i\rbrack $, $\lbrack H_i,A_{i+1}\rbrack $ such that $\tau_{\lbrack A_i,H_i\rbrack}+\tau_{\lbrack H_i,A_{i+1}\rbrack}=-\delta \tau_i$. For simplicity, and again without loss of generality, we show the existence of $H_i$ on the $1+1$ spacetime subspace identified by the segment  $\lbrack A_i,A_{i+1}\rbrack$ and by the $t$-axis.

We first select a time value $\delta \tilde t$ such that $0<\delta \tilde t<\delta t_i$, and we build the line of simultaneity $\mathcal{L}$ such that $t=\delta \tilde t$. We choose $\delta \tilde t$ small enough that the intersection point $\bar H$ of $\mathcal{L}$ and the light cone of $A_i$ makes the segment $\lbrack \bar H,A_{i+1}\rbrack$ time-like. Under these conditions, $\tau_{\lbrack A_i,\bar H\rbrack}=0$ and $\tau_{\lbrack \bar H,A_{i+1}\rbrack}>0$, so that $\tau_{\lbrack A_i,\bar H \rbrack}+\tau_{\lbrack \bar H,A_{i+1}\rbrack}>0$.

If, on the other hand, we select $\tilde H$ on  $\mathcal{L}$ that is outside the light cone of $A_i$ and sufficiently far from $A_i$ and $A_{i+1}$, the segments $\lbrack A_i,\tilde H \rbrack$ and $\lbrack \tilde H,A_{i+1}\rbrack$ will be space-like, with large negative spacetime intervals that we can denote respectively as $\delta s_1^2$ and $\delta s_2^2$. 

Since  $0<\delta \tilde t<\delta t_i$, the time differences $\delta t_1=\delta \tilde t-0$ (the time interval in $\mathcal{F}_0 $ along $\lbrack A_i,\tilde H \rbrack$) and $\delta t_2=\delta t- \delta \tilde t$ (the corresponding time interval along 
$\lbrack \tilde H,A_{i+1}\rbrack$) will be positive. It follows that  $\tau_{\lbrack A_i,\tilde H \rbrack}=\frac{\delta s_1^2}{\delta t_1^2}\delta t_1<0$ and $\tau_{\lbrack \tilde H,A_{i+1}\rbrack}=\frac{\delta s_2^2}{\delta t_2^2}\delta t_2<0$. More particularly, $\tilde H$ can be chosen so that $\tau_{\lbrack A_i,\tilde H \rbrack}+\tau_{\lbrack \tilde H,A_{i+1}\rbrack}<<-\delta \tau_i$.

We can then find an intermediate point $H_i$ between $\bar H$ and $\tilde H$ on $\mathcal{L}$ such that $\tau_{\lbrack A_i,H_i \rbrack}+\tau_{\lbrack H_i,A_{i+1}\rbrack}=-\delta \tau_i$, as needed. 

A similar argument can be used to show that  if $\tau_{\lbrack A_i,A_{i+1}\rbrack}=-\delta \tau_i<0$, we can always build two connected segments $\lbrack A_i,H_i\rbrack $, $\lbrack H_i,A_{i+1}\rbrack $ such that their total internal time is $\delta \tau_i$.

If the segment $\lbrack A_i,A_{i+1}\rbrack $ is null, then $\tau_{\lbrack A_i,A_{i+1}\rbrack}=0$. In this case, we simply replace $\lbrack A_i,A_{i+1}\rbrack $ with the pair of segments  $\lbrack A_i,H_i\rbrack $, $\lbrack H_i,A_{i+1}\rbrack $, with $H_i$ any point in $\lbrack A_i,A_{i+1}\rbrack $, and we have two null segments with $\tau_{\lbrack A_i,H_i \rbrack}+\tau_{\lbrack H_i,A_{i+1}\rbrack}=0$.

By working on each segment $\lbrack A_i,A_{i+1}\rbrack $ we build this way a new piecewise path $\gamma'_{(AB,2N-1)}$ of $2N-1$ segments denoted by the sequence of points $\{A_1,H_1,A_2,H_2,\dots,H_{N-1},A_N\}$ and such that if $\tau_{\gamma_{(AB,N)}}=\tau_N$, then  $\tau_{\gamma'_{(AB,2N-1)}}=-\tau_N$.  

Denote by $\Gamma_{(AB,N)}$ the space of piecewise linear paths made of $N$ segments, and define the map $\Sigma_N: \Gamma_{AB,N}\rightarrow \Gamma_{(AB,2N-1)}$ such that $\gamma'_{(AB,2N-1)}$ is the image of $\gamma_{(AB,N)}$. 

In the limit of $N\rightarrow \infty$, $\gamma_{AB}$ is the limit of $\gamma_{(AB,N)}$. Let $\Sigma$ be the limit of $\Sigma_N$, and note that $\Gamma_{(AB,N)}$, $\Gamma_{(AB,2N-1)}$ both converge to the space of (Brownian) indistinguishable paths $\Gamma_{AB}$.

Then $\Sigma_N(\gamma_{(AB,N)})$ will also converge point-wise to the path $\gamma_{AB}$  (even though its stochastic structure will be different in terms of distribution of time-like, space-like and null local approximations). Since this is true for every path, and distances among paths are respected in the limit, we expect the measure $D\Sigma(\gamma_{AB})$ to be the same as $D\gamma_{AB}$.\qed

\bigskip
\noindent
We conclude that the choice of total internal time along a  path given in Definition~\ref{tempogiusto} is compatible with the experience of simple units, and also with the time invertibility Condition~\ref{timeInv} .

\begin{remark}
\label{old3_7}
Clearly, if $\tau=\int_{\gamma_{AB}} \frac{1}{c^2}\frac{ds}{dt}*\frac{ds}{dt}dt$ satisfies the time invertibility Condition 1, $-\tau$ will also satisfy it. We will use this simple observation again in Section 4 while deriving the standard path integral formulation of quantum mechanics for spinless elementary particles from the general framework of internal time path integrals.
 \end{remark}
 \begin{remark}
\label{old3_8}
In the definition $d\tau=\frac{1}{c^2}\frac{ds}{dt}*\frac{ds}{dt}dt$, we could replace $L=\frac{1}{c^2}\frac{ds}{dt}*\frac{ds}{dt}$ with any polynomial in $L$, and $d\tau$ would still be compatible, in principle, with the general setting of units of experience. However, polynomials in $L$ of degree $>1$ would not lead to a relativistic invariant  $\mathcal{P}(\tau_{\gamma_{AB}})$, as it can be easily seen by replacing $L$ in  Equation~\eqref{eq:calc_relative} with a polynomial in $L$ of higher degree.
\end{remark}

\begin{remark}
\label{alpha}
There is a crucial consequence of taking $d\tau=\frac{1}{c^2}\frac{ds}{dt}*\frac{ds}{dt}dt$ as the correct expression for the internal time differential of isolated units. Since all measurable quantities are derived experimentally in the context of a time differential $d\tau=(ds^2/c^2)^{1/2}$, it follows that the value of the internal frequency $\omega_0$ to be used in the internal time path integral cannot be measured directly from experimental evaluations. Internal frequency at rest must be defined with respect to internal time differential $d\tau=\frac{1}{c^2}\frac{ds}{dt}*\frac{ds}{dt}dt$. To this extent, in Definition~\ref{innerclock} we allowed $\omega_0$ to be a multiple $\alpha$ of the de Broglie frequency. We will show in Section 4 that consistency with standard quantum mechanics in the limit of weak fields and low velocities implies $\alpha=\frac{1}{2}$. 
\end{remark}

The connection between the propagation of simple isolated units and quantum mechanics is fully fleshed out in Section 4, where we shall assume that only low energy classical paths are associated to the propagation of a unit, so that higher order relativistic effects can be neglected and the internal clock frequency at rest $\omega_0$ can be used as an approximation of $\omega_u$ in all calculations.

\section{The Propagator's Path Integral for Spinless Elementary Particles}

Let us assume now that a simple isolated unit $S$ is subject to a scalar potential $\phi(x)$. The principle of isolated equivalence implies that, irrespective of the nature of the potential, the effects of $\phi(x)$ on the internal time state along indistinguishable paths of $S$ are the same of those of a corresponding scalar gravitational potential $\Phi$, what we call the {\it induced potential} as experienced by the isolated unit. 

However, when we represent a general potential as an induced potential, we need to scale it appropriately by the mass of the unit subject to the potential. For example, in the case of a charged simple unit $S$ of charge $e$ and mass $m_e$ subject to a static electric potential $\phi$, the corresponding potential energy is $V(x)=e\phi$, we define then the induced potential experienced by the unit in isolation as  
$\Phi=\frac{e\phi}{m_e}$, so that $V(x)=\Phi m_e$ as it would be the case for a gravitational potential.

Given that for for isolated simple units there is this simple correspondence between a general, non-gravitational scalar potential $\phi$ and its induced potential $\Phi$, in the following we can work directly and without loss of generality with a gravitational potential $\Phi$. 

We further restrict ourselves in this Section to the case of a static, radially symmetric gravitational potential $\Phi$ that is weak, that is, such that $\Phi<<c^2$. Under this last assumption, and following closely \cite[Section~17.9]{GE1}, we can approximate the line element $\frac{ds^2}{c^2}$ as:
\begin{equation}
\frac{ds^2}{c^2}=(1+\frac{2\Phi}{c^2})dt^2-(1-\frac{2\Phi}{c^2})\frac{1}{c^2}d\sigma^2
\end{equation}
where $d\sigma^2=dx^2+dy^2+dz^2$. This expression for the line element provides the necessary information on the local metric to be able to compute  the Lorentzian inner product in 
$d\tau=\frac{1}{c^2}\frac{ds}{dt}*\frac{ds}{dt}dt$.
In particular, the Lorentzian inner product of two vectors $P_1=(t_1,x_1,y_1,z_1)$ and $P_2=(t_2,x_2,y_2,z_2)$ will be defined locally by
$P_1*P_2=g_{11}t_1t_2+g_{22}x_1x_2+g_{33}y_1y_2+g_{44}z_1z_2$ where 
$(g_{11},g_{22},g_{33},g_{44})=(1+\frac{2\Phi}{c^2},-(1-\frac{2\Phi}{c^2}),-(1-\frac{2\Phi}{c^2}),-(1-\frac{2\Phi}{c^2}))$.

Note that the principle of isolated equivalence ensures that, for a simple isolated unit, a unique local {\it induced metric} can always be defined for any scalar potential, simply by establishing the corresponding induced potential at that point.

The total internal time along an indistinguishable path from a point $A$ to a point $B$ is:
\begin{equation}
\label{eq:tot_time}
\begin{split}
\tau_{\gamma_{AB}}=\int_{\gamma_{AB}}\frac{1}{c^2}\frac{ds}{dt}*\frac{ds}{dt}dt =
\int_{ \gamma_{AB}} \lbrack (1+\frac{2\Phi}{c^2})-(1-\frac{2\Phi}{c^2})\frac{v^2}{c^2} \rbrack dt
\end{split}
\end{equation}
where we substituted $v^2=(\frac{dx}{dt})^2+(\frac{dy}{dt})^2+(\frac{dz}{dt})^2$, and we assume that $t$ is measured in an external frame of reference $\mathcal{F}_A$ at rest with respect to the potential field.

If $v<<c$, then $\tau_{\gamma_{AB}}$ can be approximated as:
\begin{equation}
\tau_{\gamma_{AB}}=\int_{ \gamma_{AB} } (1+\frac{2\Phi}{c^2}-\frac{v^2}{c^2}) dt
\end{equation}
in which case the internal time path integral is approximately
\begin{equation}
I(A,B)=\int e^{i\omega_u \int_{\gamma_{AB}} (1+\frac{2\Phi}{c^2}-\frac{v^2}{c^2}) dt}D\gamma_{AB}.
\end{equation}

Since we assume a non-relativistic setting, the speed $u$ of the simple unit at $A$ is such that $u<<c$, and we can approximate $\omega_u\approx\omega_0$.
Define $$K=\int e^{i\omega_0 \int_{\gamma{AB}} 1dt},$$ which is independent of the path, and substitute $\omega_0=\frac{\alpha M_0c^2}{\hbar}$ from Equation~\eqref{eq:alphaomega}, then:
\begin{equation}
\label{eq:I_phi}
I(A,B)=K \int e^{i \frac{\alpha}{\hbar} \int_{\gamma_{AB}} {2\Phi}M_0-M_0{v^2} dt}D\gamma_{AB}.
\end{equation}

We set $V=\Phi M_0$ and divide and multiply by $2$ to write
\begin{equation}
I(A,B)=K \int e^{i \frac{\alpha}{\hbar/2} \int_{\gamma_{AB}} V(x)-\frac{1}{2}M_0{v^2} dt}D\gamma_{AB}.
\end{equation}

Recalling the time symmetry argument in Remark \ref{old3_7}, the sign of the integrand in $$ \int_{\gamma_{AB}} V(x)-\frac{1}{2}M_0{v^2} dt$$ does not affect the value of $P(A,B)=|I(A,B)|^2$, and we can conclude that: 
\begin{equation}
\begin{split}
P(A,B)=|I(A,B)|^2=
| \int e^{i \frac{\alpha}{\hbar/2} \int_{\gamma_{AB}} -V(x)+\frac{1}{2}M_0{v^2} dt}D\gamma_{AB}|^2,
\end{split}
\end{equation}
where, since we deal with a probability distribution, we have absorbed the constant $K$ into the measure $D\gamma_{AB}$.

We recall now that the probability distribution $P_{qm}(A,B)$ of a quantum spinless point particle with Lagrangian  $-V(x)+\frac{1}{2} v^2 M_0$ and propagating from $A$ to $B$ can be described as follows \cite{Feynman2}:
\begin{equation}
P_{qm}(A,B)=|\int e^{i \frac{1}{\hbar} \int_{\gamma_{AB}} -V(x)+\frac{1}{2}M_0{v^2} dt}D\gamma_{AB} |^2.
\end{equation}

There is an obvious difference between $P(A,B)$ and $P_{qm}(A,B)$: in $P(A,B)$, the constant $\hbar$ is replaced by $\hbar/2$. The only way to make $P(A,B)$ consistent with quantum mechanics in the limit of weak potentials and low velocities is to set $\alpha=\frac{1}{2}$ in $\omega_0=\frac{\alpha M_0c^2}{\hbar}$ in Definition~\ref{innerclock}. That is, if we compute physical quantities by using the standard definition of time and therefore we use the probability density $P_{qm}(A,B)$, the observed internal frequency of spinless elementary particles will be close to twice the internal frequency to be used in $P(A,B)$.

In the limit of weak potentials and velocity $v<<c$, we conclude that the square module of the internal time path integral for a simple isolated unit can be reduced to the square module of the path integral describing  the propagator of quantum spinless point particles.

\begin{remark}
\label{old4_1}
In moving from the internal time path integral to its approximation, we assume that we can set $v<<c$ in the argument of the path integral while keeping the measure $D\gamma_{AB}$ unchanged. This is essentially equivalent to a regularization of the integral argument, and to the assumption that the integral is well approximated under such regularization. 
We make the same approximation in Section 5 when we calculate possible effects of a weak Coulomb potential on the local metric defined along each indistinguishable paths.
\end{remark}

The arguments derived in this Section for scalar potentials can be adapted in a straightforward way to more general vector potentials, if we continue to restrict ourselves to simple units with no spin\footnote{We note however that the theory extends, at least formally, to the more complex case of units with spin. The discreteness of spin states preserves the possibility of having a suitable space of continuous indistinguishable paths on which to build an internal time path integral.}. More particularly, for a spinless elementary particles of rest mass $m_e$ in a non relativistic regimen subject to a general electromagnetic field, we note that $V(x)$ along a path $x(t)$ takes the form $V(t)=-e\phi+\frac{e}{c} \boldsymbol A \cdot \dot x$, where $\phi$ is the scalar potential, $\boldsymbol A$ is the vector field associated to the external charges distribution and motion, $e$ is the charge of the simple unit, and we denote by $\dot x$ the derivative of $x$ along the path and by ``$\cdot$'' the Euclidean inner product.

Because the internal state variables of the simple unit cannot change in time, paths subject to vector and scalar potentials are not distinguishable for simple units. At each instant $t$, the simple unit behaves as if it is subject to a scalar potential $\Phi$ such that $V(t)=-e\phi+\frac{e}{c} \boldsymbol A \cdot \dot x=\Phi m_e$. The computation of $I(A,B)$ in Equation~\eqref{eq:I_phi} is modified (in the non-relativistic regimen) by setting the induced potential as $\Phi=({-e\phi+\frac{e}{c} \boldsymbol A \cdot \dot x})/{m_e}$ and allowing a dependence of $\Phi$ on $t$ in the path integral. 

The possibility of this straightforward generalization to vector potentials also implies that the range of indistinguishable paths of a simple unit in a scalar potential is the same as the range of indistinguishable paths in a general potential. On the other hand, the spacetime metric induced by the vector potential at each point is not uniquely defined: the metric of the line element is dependent not just on the specific test unit, but also on the indistinguishable path of the unit.

\begin{remark}
In \cite{mueller} the propagator of a quantum mechanical particle was also derived from de Broglie internal clock hypothesis in the context of weak gravitational potentials, while in \cite{dolce1,dolce2} the Feynman path integral propagator in the presence of general gauge fields is derived from spacetime geometrodynamics. Another geometric interpretation of quantum mechanics valid for general potentials was proposed in \cite{tavernelli}, starting from Bohm's pilot wave interpretation of quantum mechanics.

In our work the effects on the local metric induced by a general potential are deduced from first principles; the framework of isolated units and the principle of isolated equivalence allow to extend the analysis of weak gravitational potentials to scalar and vector potentials for simple units. Moreover, our framework is not an interpretation of quantum mechanics, as it reduces to the latter only under the condition of weak potentials and low velocities, when the propagator of a simple unit in Eq.~\eqref{eq:Wgamma} approximates the quantum mechanical propagator of a spinless elementary particle\footnote{As noted in Remark~\ref{radiation}, for a charged unit in a general electromagnetic potential the predictions of the standard theory are preserved under most experimental conditions because of the intermittent loss of isolated status of the unit due to its radiating field.}. We shall see Section 5 that the condition of isolated evolution of a unit can be harnessed to device simple experimental tests of our theory.
\end{remark}

\section{Weak Relativistic Effects and Experimental Considerations}

We now explore whether it is possible to experimentally detect induced effects due to a non-gravitational potential on the local spacetime metric defined along the indistinguishable paths of a simple unit. 

The whole experimental setting we suggest here is meaningful only under restrictive conditions on the potential and the extremal trajectories of charged simple units. These conditions are meant to ensure that: the radiating field of the unit is negligible and does not impact its isolated status; and that the results of Section 4 can be extended to include weak relativistic effects of the order of $\frac{u^2}{c^2}$, with $u$ the velocity of the unit at the beginning of propagation.

We avoid any scenario that would require an accurate evaluation of $P(A,B)=|I(A,B)|^2$, because $d\tau= \frac{1}{c^2}\frac{ds}{dt}*\frac{ds}{dt}dt$ defines indirectly a local Lorentzian geometry, but in general the internal time path integral is not defined on a single Lorentzian manifold. A potential that depends on the velocity of the unit (such as, for example, a vector potential) will induce on each indistinguishable path its own local metric. A proper perturbation theory for such type of path integrals is not available.

In light of these computational limitations, in the following we work only with static potentials and on measurements of the shape of extremal paths associated to the internal time path integral that are not essentially dependent on a complete perturbative analysis of integral itself.

\paragraph{Weak Relativistic Effects.} Our derivation of  quantum mechanics for spinless elementary particles in Section 4 assumes a nonrelativistic regimen where the relativistic expression $\omega_u$ for the internal clock frequency was replaced by the approximation $\omega_0$. However, the most general expression for the internal time path integral in Equation~\eqref{eq:innerInt} used $\omega_u$ and was relativistically invariant for the frames of reference defined at the matching points. This raises the possibility that our formalism could be extrapolated to consider weak relativistic effects, while eschewing the difficulties of a fully relativistic setting.

Accordingly,  in this Section we continue to assume that $v<<c$, but we seek very small effects of the order of $\frac{v^2}{c^2}$, and in particular of the order of $\frac{u^2}{c^2}$, where $u$ is the velocity of the simple unit as established at the point $A$ in an external frame of reference at rest with respect to the potential field and with the whole experimental apparatus. We further assume that we are in the presence of static, radially symmetric Coulomb potential $\phi$, and that the induced potential $\Phi$ due to the Coulomb potential is weak, i.e. $\Phi<<c^2$. 

For an electron of charge $e$ and rest mass $m_e$, the relativistic internal clock frequency is $\omega_u=\epsilon\frac{\alpha m_ec^2}{\hbar}$, with $\alpha=\frac{1}{2}$ (as we established in Section 4), and $\epsilon=(1-\frac{u^2}{c^2})^{-1/2}$. Similarly, if we wish to consider weak relativistic effects, the relativistic mass $\epsilon m_e$ will appear in the computation of the induced potential due to the static electric potential, as $\Phi=\phi \frac{e}{\epsilon m_e}$.

Following the expression for the total internal time along a path given in Equation~\eqref{eq:tot_time}, the corresponding approximate internal time path integral for the induced weak potential is: 
\begin{equation}
\label{eq:columb_path}
I(A,B)=\int e^{i\omega_u \int_{\gamma_{AB}} \lbrack (1+\frac{2\Phi}{c^2})-(1-\frac{2\Phi}{c^2})\frac{v^2}{c^2} \rbrack dt}D\gamma_{AB}.
\end{equation}

We can assume that there is only one extremal path from $A$ to $B$ for $I(A,B)$. Multiple extremal paths among points $A$ and $B$ may arise for strong potentials. However the energy of the particle along extremal paths that do not correspond to the classical trajectory between $A$ and $B$ would generally be high. In these cases relativistic effects would be significant, while we are considering only the case of paths with initial velocity $u<<c$ and subject to weak potentials.

The relativistic mass affects the extremal trajectories of the path integral in Equation~\eqref{eq:columb_path} through the equality $\Phi=\phi \frac{e}{\epsilon m_e}$. The relativistic internal clock frequency $\omega_u$ is also dependent on the relativistic mass, but the specific value of $\omega_u$ affects only the value of the path integral, and not its extremal trajectories.

\begin {remark}
\label{radiation}
The integral in Equation~\eqref{eq:columb_path} has extremal trajectories that differ in general from the classical trajectory of a charged particle in an static electric potential, even in the limiting case of weak potentials. The impact of the relativistic mass is fixed at the point of matching of experiences in $A$ and does not change as long as the unit is isolated. This is a consequence of the principle of isolated equivalence that we need to reconcile with the predictions of the classical theory of electromagnetism, where the equations of motion depend on the relativistic mass.

To this end, we note that the electromagnetic radiation of the electron is unavoidable in the context of any significant acceleration. This radiation would affect the state of the surrounding environment and it would make the unit, at least intermittently, not isolated. The intermittent matching of experiences would achieve two results: break the isolated equivalence that gives rise to induced metric effects; and reset every time the value of $u$ in Equation~\eqref{eq:columb_path}. In this way a correspondence with the standard classical trajectory of a charge accelerated in an electromagnetic field is preserved. Note however that a full account of radiating interaction with other units can be properly addressed only by generalizing the current work to include structured units.
\end{remark}

\paragraph{Experimental Considerations.} The experimental setting we assume in this Section is essentially a single slit diffraction experiment (see for example \cite[Chapter 3.2]{Feynman2}), where we pay particular attention to the distance of screen and detector and where we add a radially symmetric weak Coulomb potential between screen and detector. 

Let  $x_a$ be the position of the mouth of an electron gun pointing right at a screen that is orthogonal to the line $l$ that goes through the gun mouth. Set a slit on the screen at the intersection of the screen and the line $l$. Position a detector plane parallel to the screen and to its right. Suppose that electrons are emitted with initial velocity that is very narrowly distributed around some value $u$, so that without loss of generality we can consider only paths with initial velocity exactly $u$. Assume also that a radially symmetric Coulomb potential is centered at a distance $D$ from the line $l$ and to the right of the screen. Let $A=(x_a, t_a)$ with $x_a$ at the mouth of the electron gun, and $B=(x_b,t_b)$  with $x_b$ a point on the detector and $t_b>t_a$.

The velocity of the electron is assumed to be large enough that the potential field only deflects the electron and it does not significantly change the norm of the velocity. These assumptions are crucial also to ensure that the transversal acceleration is small, so that the relative loss of kinetic energy due to electromagnetic radiation can be neglected as well. We stress that a very low radiating field is necessary to ensure the electron will be an isolated unit until it reaches the detector. 

We first analyze a particular path in this experimental setting, the one corresponding to the classical trajectory with initial velocity $u$. We then argue that that the maximum of the distribution of the electron's hits on the detector is located at the intersection of this path with the detector itself.

Note that since the velocity along the extremal trajectory is assumed to be nearly constant, the relativistic mass is constant as well.  Under this condition, the caveats of Remark~\ref{radiation} do not apply, and the computation of the deflection of the extremal path associated to the path integral in Equation~\eqref{eq:columb_path}, with starting point $x_a$ at the mouth of the electron gun  and velocity equal to $u$ follows the calculations for the deflection of finite mass particles in a weak gravitational potential. 

In a radial, static gravitational potential that is flat at infinity, the total angle deflection of the trajectory of a particle of initial velocity $u$ can be approximated as \cite[Section~25.5]{Misner}:
\begin{equation}
\label{eq:misner}
\delta \alpha_G \approx 2\frac{GM}{D}\frac{1}{u^2}(1+\frac{u^2}{c^2})
\end{equation}
$D$ is the distance of the closest point of the trajectory to the potential field source (the impact parameter), $M$ is the total mass of the field source, and we assume that $u$ is large enough to allow the particle to escape the field. 

The scalar, induced potential for a given static Coulomb potential is, in radial coordinates, 
\begin{equation}
\label{eq:phiC}
\begin{split}
\Phi(R)=k_e\frac{Qe}{ R} \frac{1}{\epsilon m_e}=
k_e\frac{Qe}{ R} \frac{1}{m_e}(1-\frac{u^2}{c^2})^{1/2},
\end{split}
\end{equation}
with $k_e$ the Coulomb constant, $Q$ the charge of the Coulomb field, $e$ the charge of the electron, and $m_e$ its mass.  

Let $\Gamma$ be the unique geodesic (of the geometry induced by the potential $\Phi$ in Equation~\eqref{eq:phiC}) from $x_a$ to the detector plane with initial velocity $u$. Substituting $\Phi(D)$ for $\frac{GM}{D}$ in Equation~\eqref{eq:misner} we have this corresponding formula for the deflection angle of $\Gamma$:

\begin{equation}
\delta \alpha \approx 2 k_e\frac{Qe}{D}\frac{1}{m_e}(1-\frac{u^2}{c^2})^{1/2}\frac{1}{u^2}(1+\frac{u^2}{c^2}).
\end{equation}

Let now $B_{\Gamma}$ be the intersection of $\Gamma$ and the detector plane, and denote by $P_A(x_b,\Phi)$ the restriction of the probability distribution $P(A,B)$ on the detector plane in the presence of the potential $\Phi$. 

Since $\Gamma$ is the only permissible classical path in our experimental setting, it is the only extremal path that will contribute to the corresponding WKB semiclassical limit. In this limit, it will be the path reaching the detector with the largest constructive interference of nearby paths. Accordingly, the maximum of $P_A(x_b, \Phi)$ is located at $B_{\Gamma}$ and the deflection angle $\delta \alpha$ is measurable from $P_A(x_b, \Phi)$.

In a classical setting, without induced effects on the local metric, the deflection angle due to the Coulomb field would be \cite[page~671]{Misner}:
\begin{equation}
\delta \alpha_C \approx 2k_e\frac{Qe}{D}\frac{1}{m_e}(1-\frac{u^2}{c^2})^{1/2}\frac{1}{u^2}.
\end{equation}

This calculation considers effects of first order in $\frac{u^2}{c^2}$ and strictly speaking we cannot apply standard non-relativistic quantum mechanics here. However, since $u<<c$ and since in the WKB limit we expect to recover the classical trajectory also for relativistic quantum mechanics, we can assume that the deflection angle $\delta \alpha_C$ will correspond to the peak of the distribution of hits on the detector plane as predicted by a standard quantum mechanical analysis of this setting.

By approximating $(1-\frac{u^2}{c^2})^{1/2}=1-\frac{1}{2}\frac{u^2}{c^2}$ and retaining only terms of the order of $\frac{u^2}{c^2}$ we can write $\delta \alpha-\delta \alpha_C\approx 2k_e \frac{Qe}{Dm_ec^2}$. This difference will be small in absolute terms, because the condition of weak induced potential $\Phi<<c^2$ implies that $2k_e \frac{Qe}{Dm_ec^2}<<1$. Still, $\delta \alpha-\delta \alpha_C$ gives an estimate of the interval of accuracy for the measurement of  $\delta \alpha$ that would be required to validate experimentally the weak relativistic effects conjectured in this section.

We conclude that the maximum of the probability distribution of hits on the detector plane should be deflected by the Coulomb field by an angle $\delta \alpha$ and not by $\delta \alpha_C$, as it would be predicted by standard quantum mechanics. Once more, we stress this result would hold only assuming the radiating field of the electron can be neglected and does not affect its isolated status, and extrapolating the results of Section 4 to hold when effects of the order $\frac{u^2}{c^2}$ are very small, but not negligible.

\section{Conclusion}

This work started from the assumption that the evolution of a physical phenomenon (of any size) is subordinate to the restricted information available to the phenomenon itself through its state variables (the experience of the unit), whenever no other phenomenon validates this information. From this broad assumption, we developed a primitive ontology of isolated units, i.e.  phenomena that do not affect significantly their surrounding environment. Such units were then assumed to intermittently match their experiences during interaction. 

The main physical implications of the primitive ontology of isolated units are the following: the equivalence principle can be conditionally extended to non-gravitational potentials whenever an isolated unit cannot distinguish the nature of the potentials via its state variables; similarly, all spacetime paths of an isolated unit that cannot be distinguished must be considered physically real in isolation, and must eventually have an impact when the unit interacts with other units. 

From these assumptions, we derived a relativistically invariant propagator for simple units (isolated units whose state variables do not change), under an additional assumption on the existence of an internal periodic representation of time for nonzero mass units (adapted from de Broglie's internal clock frequency hypothesis). We then demonstrated that the propagator for simple units reduces to the standard quantum mechanical propagator of spinless elementary particles in the limiting case of weak potentials and low velocities. By extrapolating these results to include weak relativistic effects, and under restrictive conditions meant to minimize field radiation and ensure that a moving electron can be considered an isolated unit, we suggested an experimental setting that could detect small deviations from the predictions of standard quantum mechanics for the propagator of an electron slowly moving in a weak Coulomb potential.

In fact, as pointed out in Remark~\ref{radiation}, radiating charged particles only intermittently retain isolated status, and the framework of simple units of experience is insufficient to account for a fully relativistic theory of moving charged units, and it will need to be extended to include structured units. This extension will be necessary also for a proper treatment of units with spin. 

Note however that the isolation of a unit is a non-local, relational property,  and in principle it is amenable to experimental manipulation, by controlling its surrounding environment. We expect therefore that an experimental and phenomenological exploration of the notions of isolated unit and of the matching of experiences will yield further interesting results, even before a complete theory for structured units is developed. 

In particular, we expect the range of permissible indistinguishable paths of large molecules to be related to the topology of their molecular configurations, and to the way these configurations respond to the induced tidal effects of general potentials. A deeper understanding of this relation may allow to establish qualitative constraints on the observed quantum superposition \cite{arndt} that persists for large, composite objects,  and contribute a different viewpoint on the transition from quantum to classical behavior.

%


\end{document}